\documentclass[final,5p,times,twocolumn]{elsarticle}

\usepackage{amssymb}
\usepackage{amsmath}
\usepackage{xcolor}
\usepackage[normalem]{ulem}

\journal{Physics Letters B}

\begin{document}

\begin{frontmatter}
\title{Bottomonium production in an open quantum system approach\\ with interactions from lattice quantum chromodynamics}

\author[a]{Shuhan Zheng}
\ead{Corresponding author: zhengsh24@mails.tsinghua.edu.cn}
\author[a,b]{Shuzhe Shi}
\ead{shuzhe-shi@tsinghua.edu.cn}
\affiliation[a]{organization={Department of Physics},
            addressline={Tsinghua University}, 
            city={Beijing},
            postcode={100084}, 
            country={China}}
\affiliation[b]{organization={State Key Laboratory of Low-Dimensional Quantum Physics},
            addressline={Tsinghua University}, 
            city={Beijing},
            postcode={100084}, 
            country={China}}

\begin{abstract}
Bottomonium production in Pb-Pb collisions at $\sqrt{s_{NN}}=5.02$ TeV is studied using a Lindblad master equation derived from potential non-relativistic quantum chromodynamics (QCD), where quantum regeneration of color-singlet states is matched to the lattice QCD imaginary potential via collapse operators. Two parametrizations of the in-medium heavy-quark potential, both constrained by lattice QCD data, are employed to compute the nuclear modification factors of $\Upsilon(1S)$, $\Upsilon(2S)$, and $\Upsilon(3S)$. The results show sensitivities to both the quantum regeneration effect and the initial condition of the density matrix. The dipole transitions in the collapse operators are found to significantly redistribute populations among different orbital angular momentum channels.
It is shown that regeneration is more important when a potential with a larger imaginary part, i.e., stronger transitions between singlet and octet states, is used.
\end{abstract}



\begin{keyword}
relativistic heavy-ion collisions \sep quark-gluon plasma \sep heavy flavor \sep open quantum system



\end{keyword}

\end{frontmatter}




\section{Introduction}
\label{introduction}

Ultrarelativistic heavy-ion collisions at the LHC and RHIC create a deconfined state of strongly interacting matter---the quark-gluon plasma (QGP)---providing a unique laboratory for studying QCD under extreme conditions. Heavy quarkonium, the bound state of a heavy quark-antiquark pair ($c\bar{c}$ or $b\bar{b}$), serves as a sensitive probe of the QGP: since the seminal work of Matsui and Satz~\cite{Matsui:1986dk}, its suppression in heavy-ion collisions has been recognized as a signature of deconfinement, and extensive subsequent studies~\cite{Blaizot:1996nq, Zhu:2004nw, Yan:2006ve, Bratkovskaya:2004ec, Braun-Munzinger:2000csl, Grandchamp:2003uw, Andronic:2015wma} have established that both color screening and Landau damping---encoded in the real and imaginary parts of the in-medium heavy-quark potential, respectively---govern quarkonium dissociation in the hot medium.

For heavy quarkonium, the inherent hierarchy of energy scales---the heavy quark mass $M$, the relative momentum $Mv$, and the binding energy $Mv^2$---enables a systematic effective field theory treatment using potential non-relativistic QCD (pNRQCD)~\cite{Brambilla:2004jw}. Within this framework, the in-medium evolution of quarkonium has been predominantly studied by solving the time-dependent Schr\"odinger equation with a complex-valued heavy-quark potential~\cite{Islam:2020gdv, Islam:2020bnp, Wen:2022yjx, Wen:2022utn}. While these Schr\"odinger-based approaches have successfully described the nuclear modification factor $R_{AA}$ and elliptic flow $v_2$ of bottomonium states, they are inherently restricted to pure-state evolution in a fixed color channel, and the imaginary potential only effectively accounts for the color-singlet to color-octet transition, but not the inverse process. Note that a Schr\"odinger-based study~\cite{Chen:2024iil} observes that implementation of the state-of-the-art lattice QCD calculations of in-medium bottomonium potentials~\cite{Burnier:2015tda, Bala:2021fkm} would lead to significant underdescription of bottomonium production in LHC energy nucleus-nucleus collisions, which was later confirmed in a data-driven Bayesian analysis of the potential~\cite{Zheng:2025bdc}. It, therefore, would be interesting to investigate the role of octet-to-singlet regeneration, which is argued to be indispensable for describing bound states in studies with phenomenological estimations~\cite{Wu:2025lcj} and the open quantum system framework~\cite{Brambilla:2023hkw, Brambilla:2024tqg}. 

The open quantum system framework provides a theoretically unambiguous description of the quarkonium quantum state taking into account the in-medium interaction with the QGP including both dissociation and regeneration through singlet-octet transitions~\cite{Brambilla:2023hkw, Brambilla:2024tqg, Yao:2020xzw, Brambilla:2022ynh}. See \cite{Yao:2021lus, Akamatsu:2020ypb} for recent reviews.
Note that in these studies, the interaction potential and operators were taken from perturbative calculations.
Motivated by the importance of regeneration production of bottomonium states, we study in the present work the effect of octet-to-singlet regeneration within the open quantum system framework by establishing a direct operator-level matching between the Lindblad equation and the nonperturbative lattice QCD results of in-medium potentials~\cite{Burnier:2015tda, Bala:2021fkm}. 

This work investigates bottomonium regeneration and suppression in Pb-Pb collisions at $\sqrt{s_{NN}}=5.02$ TeV within the open quantum system framework. The Lindblad master equation formalism is presented in Sec.~\ref{Sec:1}. Results are discussed in Sec.~\ref{Sec:results}, followed by a summary in Sec.~\ref{Sec:summary}. Natural units $k_B=c=\hbar=1$ are used throughout.

\section{Lindblad Dynamics}\label{Sec:1}
At top LHC energy, the fraction of bottomed events in $p+p$ collisions can be estimated as $\sigma_{b\bar{b}} / \sigma_\mathrm{inel} = (72\,\mu\mathrm{b}) / (70\,\mathrm{mb}) \approx 10^{-3}$~\cite{LHCb:2016qpe}. Thus, in nucleus-nucleus collisions, the $b$ and $\bar{b}$ quarks are dilute, and we may focus on a pair of $b\bar{b}$ quarks as the ``system'' of interest, whereas the remainder of the hot medium serves as the thermal ``environment''~\cite{Akamatsu:2020ypb, Brambilla:2020qwo}. Under the Markovian approximation, the density matrix $\rho(t)$ of the $b\bar{b}$ pair evolves according to a Lindblad master equation of the Gorini--Kossakowski--Sudarshan--Lindblad (GKSL) form~\cite{Lindblad:1975ef, Gorini:1975nb}. In the effective field theory of pNRQCD, the $b\bar{b}$ states are classified into color-singlet ($s$) and color-octet ($o$) configurations, and the density matrix can be separated into corresponding blocks,
\begin{equation}\label{eq:rho}
    \rho(t) = \begin{pmatrix} \rho_{ss}(t) & \rho_{so}(t) \\ \rho_{os}(t) & \rho_{oo}(t) \end{pmatrix}.
\end{equation}
Here, $\rho_{ss}$ and $\rho_{oo}$ are the density matrices of the color-singlet and color-octet subsystems, respectively, and $\rho_{so}$ and $\rho_{os}$ are the crossing terms. They are all defined on the relative-coordinate Hilbert space. The evolution of the full density matrix $\rho(t)$ is governed by the Lindblad equation
\begin{equation}\label{eq:lindblad}
    \frac{\mathrm{d}\rho}{\mathrm{d}t}
    = -i[\hat{H},\rho] + \sum_{n=0}^{1}\sum_{i=1}^{3}
    \Bigl(\hat{C}_i^n \rho\, \hat{C}_i^{n\dagger} - \tfrac{1}{2}\{\hat{C}_i^{n\dagger}\hat{C}_i^n,\rho\}\Bigr),
\end{equation}
Here $\hat{H}=\mathrm{diag}(\hat{H}_s, \hat{H}_o)$ is the in-medium Hamiltonian. The effective Hamiltonian for the color-singlet (octet) consists of the kinetic term and the real part of the in-medium potential: $\hat{H}_{s(o)}=-\nabla^2/M+V_{s(o)}(r;T)$, with $M=4.62$ GeV the bottom quark mass.
The collapse operators $\hat{C}_i^n$ encode the color-exchanging interaction of the $b\bar{b}$ pair with the medium and are of two types: $\hat{C}_i^0$ mediates singlet-octet transitions, while $\hat{C}_i^1$ induces orbital angular momentum transitions within the octet subspace. $\sum_{n,i} \hat{C}_i^n\rho \hat{C}_i^{n\dagger}$ describes the quantum fluctuations, while the corresponding anti-commutator, $-\frac{1}{2}\sum_{n,i}\{\hat{C}_i^{n\dagger} \hat{C}_i^n,\rho\}$, accounts for the quantum dissipation that drives the decay of the singlet and octet states due to their interaction with the medium.

The operator $C_i^{n=1}$ accounts for quark diffusion in color-octet states. It takes the standard next-to-leading-order pNRQCD form~\cite{ Brambilla:2023hkw, Brambilla:2022ynh, Akamatsu:2020ypb}
\begin{equation}\label{eq:C1}
    \hat{C}_i^1 = \sqrt{\frac{\tilde{\kappa}T^3(N_c^2-4)}{2(N_c^2-1)}}
            \begin{pmatrix} 0 & 0 \\ 0 & 1 \end{pmatrix}
            \Bigl(\hat{r}_i + \frac{i \hat{p}_i}{2MT}\Bigr),
\end{equation}
with $\tilde{\kappa}=4.0$ the heavy-quark momentum diffusion coefficient and $T$ the local temperature. The matrix elements of $\hat{r}_i$ and $\hat{p}_i$ are evaluated as rank-1 irreducible tensor operators, with the dipole selection rules $\Delta l = \pm 1$, $\Delta m = 0, \pm 1$. The index $i=1,2,3$ runs over the three spatial directions, reflecting the local isotropy of the QGP.
The transition operator $\hat{C}^{n=0}_i$ contains only one nonvanishing component $\hat{C}^0 \equiv \hat{C}_1^0$ with $\hat{C}_2^0=\hat{C}_3^0=0$. $\hat{C}^0$ is isotropic,
\begin{equation}\label{eq:C0}
    \hat{C}^0 = \begin{pmatrix} 0 & \hat{b} \\ \hat{a} & 0 \end{pmatrix},
\end{equation}
where the $\hat{a}$ and $\hat{b}$ operators respectively describe singlet-to-octet and octet-to-singlet transitions.
They are different by a color-degeneracy-factor $\hat{b}=\hat{a}/\sqrt{N_c^2-1}$. 

The interaction potential in the effective Hamiltonian ($\hat{H}_{s}$) and the singlet-to-octet transition operator $\hat{a}$ can be directly related to the real and imaginary potentials of the pNRQCD Schr\"odinger equation for the singlet sector.
In pNRQCD, the in-medium interaction of a static $b\bar{b}$ pair is described by a complex potential $V(r;T)=V_R(r;T)-iV_I(r;T)$~\cite{Laine:2006ns, Brambilla:2008cx}, where $V_I>0$ encodes the thermal ``decay width'' that transforms a color-singlet state into a color-octet one. From the Schr\"odinger equation one may obtain the equation of motion for the singlet density matrix,
\begin{equation}\label{eq:rhos_pNRQCD}
    \frac{\mathrm{d}\rho_{ss}}{\mathrm{d}t}
    = -i \Big[-\frac{\nabla^2}{M}+V_{R},\rho_{ss}\Big] - \Big\{V_I,\rho_{ss}\Big\}.
\end{equation}

For comparison, we focus on the singlet sector of the Lindblad equation~\eqref{eq:lindblad}. Taking the limits $\rho_{so}=\rho_{os}=\rho_{oo}=0$ to turn off the octet-to-singlet feedback, its equation of motion follows
\begin{align}\label{eq:rhos_rhoo0}
\frac{\mathrm{d}\rho_{ss}}{\mathrm{d}t} 
=
-i\Big[\hat{H}_s,\rho_{ss}\Big]
-\frac{1}{2} \Big\{\hat{a}^{\dagger}\hat{a} ,\,\rho_{ss} \Big\}.
\end{align}
Matching Eq.~\eqref{eq:rhos_pNRQCD} with Eq.~\eqref{eq:rhos_rhoo0} for arbitrary $\rho_{ss}$ yields the operator identity
\begin{equation}\label{eq:matching}
    V_{s}(r;T) = V_R(r;T),
    \qquad
    \hat{a}^{\dagger}\hat{a} = 2\,V_I(r;T),
\end{equation}
and we take $\hat{a}^{\dagger} = \hat{a} = \sqrt{2\,V_I(r;T)}$. 
To avoid model dependence, we adopt two sets of pNRQCD potentials both extracted from the state-of-the-art lattice QCD results: one is Burnier, Kaczmarek, and Rothkopf's lattice QCD calculations with $2+1$ flavors of dynamical light quarks discretized with the asqtad action, with the spectral function extracted using the Bayesian reconstruction method~\cite{Burnier:2015tda}, which exhibits an obvious screening effect in the real part~\cite{Lafferty:2019jpr}; and the other is from the HotQCD collaboration's lattice NRQCD calculation of the distance-dependent imaginary-time correlation assuming Gaussian-shaped spectral functions~\cite{Bala:2021fkm}, in which the real part remains indistinguishable from the vacuum but is associated with a stronger imaginary part. See \ref{app:B} for details regarding parametrization.
In each case, the real part of the potential contains two terms, $V_s = V_{g} + V_\mathrm{conf}$. $V_{g}$ is the attractive Coulomb/Yukawa term generated by (in-medium) gluon exchange, and $V_\mathrm{conf}$ is the confinement term. For color-octet states, the former becomes repulsive and should be modified by an extra color factor, and we argue that the confinement term should vanish. Thus, we take the octet potential $V_o = -V_{g}/8$.

We perform the evolution in the basis of vacuum Cornell eigenstates, referred to as $\{|nlm\rangle\}$, solved numerically via the inverse power method~\cite{Zhao:2024ipm, 1994JCoPh.115..470C}. We truncate the principal quantum number at $n \leq N_\mathrm{max}$ and the orbital quantum number at $l \leq l_\mathrm{max}$, respectively, for singlet and octet states. Thus, $\rho_{ss}$ and $\rho_{oo}$ are $D \times D$ matrices with $D=N_\mathrm{max}\,(l_\mathrm{max}+1)^2$, where $N_\mathrm{max}=20$ and $l_\mathrm{max}=2$ are taken in this study, and all operator matrix elements are computed by radial integration against the wavefunctions $\{u_{nl}(r)\}$. The Lindblad equation is integrated using a Taylor series expansion~\cite{Gu:2024jur}, which permits arbitrary truncation order and requires fewer working matrices than Runge--Kutta schemes, while preserving Hermiticity and trace exactly.

The initial density matrix at the formation time $t_0 = 0.6\;\mathrm{fm}/c$ is a mixed state with diagonal entries in the Cornell eigenbasis weighted by the primordial production cross sections \cite{Islam:2020bnp}:
\begin{equation}
\rho_{c}(t_0) = \sum_{nlm}\frac{f_c\,\sigma_{nl}^{\rm primordial} |nlm\rangle\langle nlm|}{(2l+1)\sum_{n'l'}\sigma_{n'l'}^{\rm primordial}},
\end{equation}
where $c \in \{ss, oo\}$. $f_{ss}$ ($f_{oo}$) is the singlet (octet) fraction, and we take two scenarios: first, $f_{ss}=2/7$ and $f_{oo}=5/7$ according to the pQCD calculation of the initial color state of $Q\bar{Q}$ pairs~\cite{Fadin:1990wx, Wong:1996et}; second, the singlet-only initial state $f_{ss}=1$ and $f_{oo}=0$, which was taken in~\cite{Brambilla:2023hkw} and justified by the negligible effect of the initial octet sector.
It might be worth noting that with such an initial condition, the density matrix always takes a block-diagonal form, i.e., $\rho_{os}(t) = \rho_{so}(t) = 0$.

\begin{figure}[tb!]
	\centering 
	\includegraphics[width=0.45\textwidth]{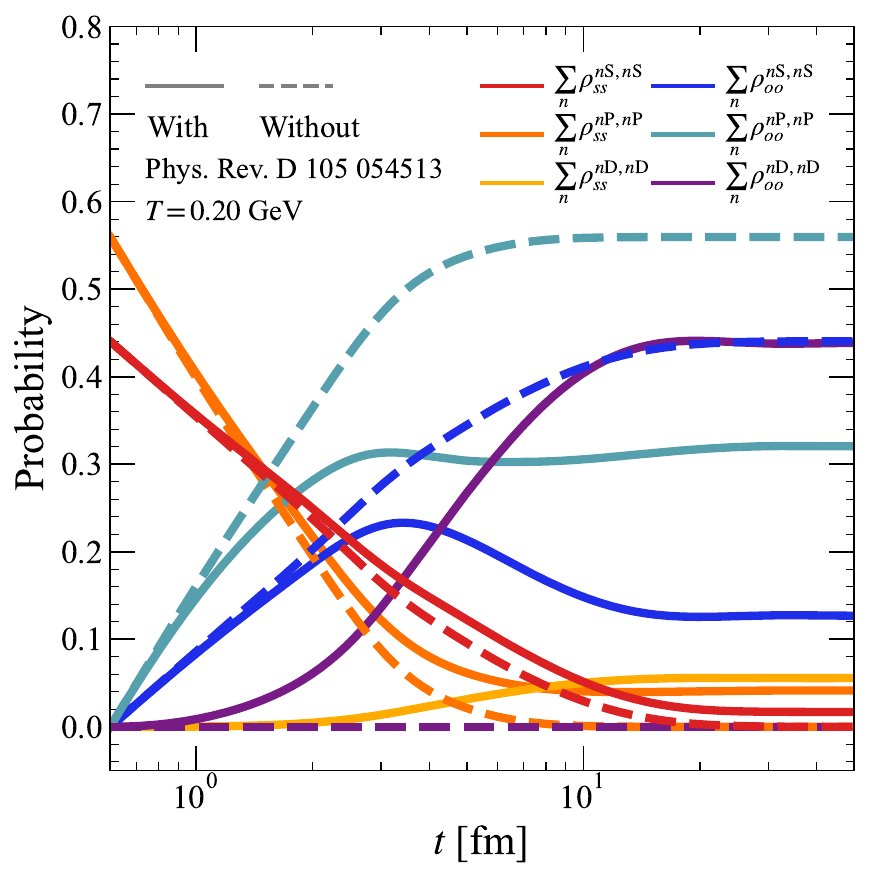}	
	\caption{(Color online) Time evolution of the diagonal density matrix elements for bottomonium in a static QGP at fixed temperature $T = 200$ MeV, evolved under the Lindblad equation with the Burnier--Kaczmarek--Rothkopf lattice QCD potential~\cite{Burnier:2015tda}.  Contributions are summed over the singlet and octet sectors for angular momentum $l = 0, 1, 2$. S-, P-, and D-waves are represented by red, orange, and gold (blue, cyan, and purple) curves, respectively, for the color singlet (octet) sector. Solid curves are for the full simulation with the singlet-only initial condition and dashed curves correspond to the comparison test that turns off the octet-to-singlet transition and octet-to-octet diffusion.}
	\label{fig:densitymatrixevolution}%
\end{figure}

With the numerical framework set up, let us begin with checking the time evolution of the density matrix elements in a static ``QGP brick'' with homogeneous and constant temperature. Figure~\ref{fig:densitymatrixevolution} shows the diagonal density matrix elements summed over the principal quantum number $n$, resolved by color subspace and orbital angular momentum. For comparison, we turn-off the octet-to-singlet feedback and octet-to-octet diffusion (by letting $\hat{b}=0$ and $\hat{C}^1_i=0$) and show the results as dashed curves. 

With the singlet-only initial condition specified above, the full simulation shares the same early-time asymptotic behavior as the simplified one without octet-to-singlet feedback or octet-to-octet diffusion. The occupation fraction of color singlet states decays exponentially and converts to octet states with the corresponding orbital and magnetic quantum numbers\footnote{Note that the singlet-to-octet conversion ($\hat{a}$) respects orbital angular momentum conservation.}. Unlike the simplified system in which the final-state density matrix contains only octet states by definition, the full evolution ends up in an equilibrium state where the singlet-versus-octet ratio approximately follows the naive color degeneracy factor $1:8$ in each orbital angular momentum (OAM) sector; meanwhile, for each color sector, ratios between different OAM states read $1:2.5:3.3$ and $1:2.5:3.5$ for singlet and octet states, respectively, which are close to, yet distinct from, the angular degeneracy $1:3:5$. The diffusion interaction, in the octet-to-octet sector, drives the quantum states to be randomly distributed in OAM quantum numbers, whereas the interaction potential in the effective Hamiltonians keeps the ratio away from that. In particular, we note that the diffusion interaction significantly enlarges the average distance between $b$ and $\bar{b}$ quarks and suppresses the recombination production proposed in~\cite{Wu:2025lcj}. 
While Fig.~\ref{fig:densitymatrixevolution} focuses on the medium at $T=0.2\;\mathrm{GeV}$ with the Burnier--Kaczmarek--Rothkopf potential that corresponds to an equilibration time $\tau \sim 10~\mathrm{fm}$, we have checked that increasing the temperature to $T=0.4\;\mathrm{GeV}$ or switching to the screening-less potential would lead to quantitatively similar final states and only speed up the equilibration by a factor of $\lesssim 2$.

\section{Bottomonium production}\label{Sec:results}
Now we move on to compute the experimental signals of bottomonium production in realistic heavy-ion collisions. We compute the nuclear modification factors ($R_{AA}$) of $\Upsilon(1S)$, $\Upsilon(2S)$, and $\Upsilon(3S)$ states in Pb--Pb collisions at $\sqrt{s_{NN}}=5.02$ TeV.

\begin{figure}[tb!]
	\centering 
	\includegraphics[width=0.45\textwidth]{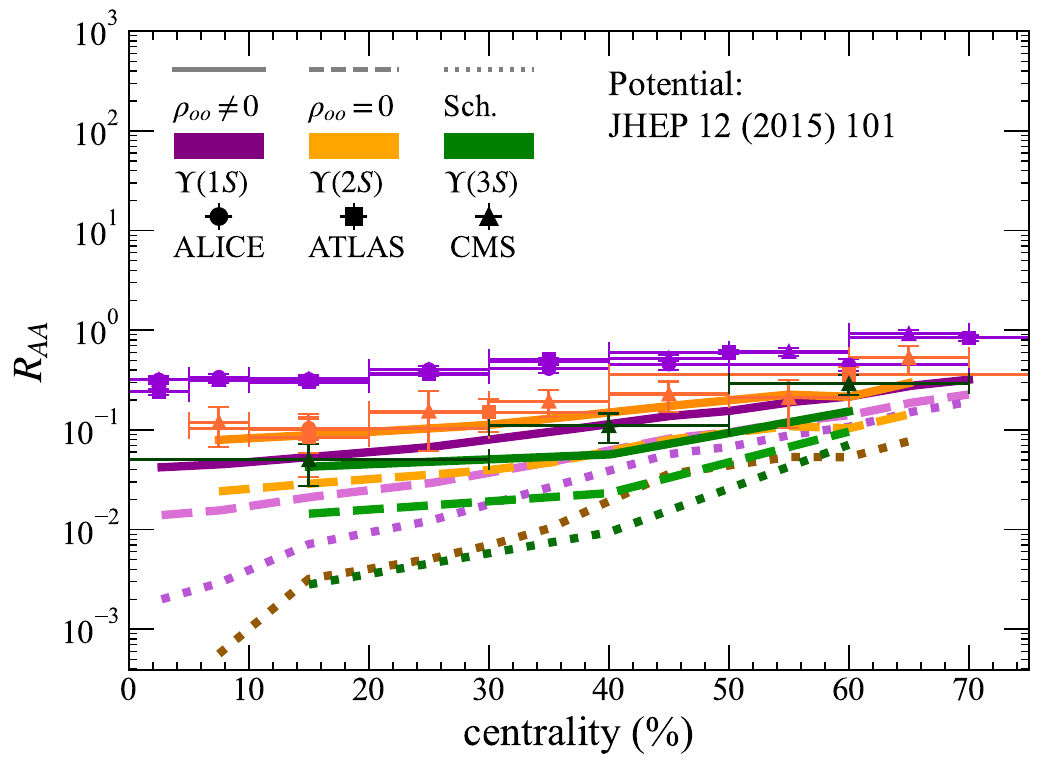}
    \includegraphics[width=0.45\textwidth]{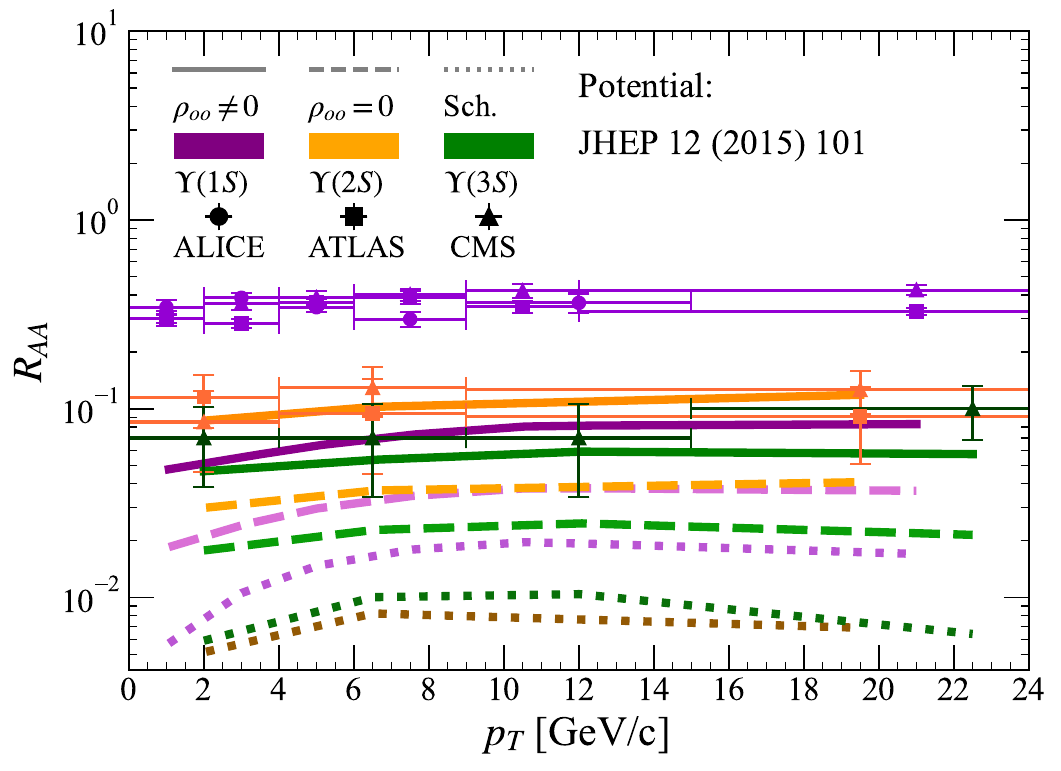}
	\caption{(Color online) Nuclear modification factors of $1S$ (purple), $2\mathrm{S}$ (orange) and $3\mathrm{S}$ (green) states of $\Upsilon$ as functions of centrality (top) and the transverse momentum (bottom), evolved under the Lindblad equation with the HotQCD potential~\cite{Bala:2021fkm}. Solid (dashed) lines correspond to evolution with pQCD motivated (singlet-only) initial conditions, whereas dotted curves are from the Schr\"odinger-based evolution, i.e., without octet-to-singlet feedback. Experimental data from the ALICE~\cite{ALICE:2018wzm}, ATLAS~\cite{ATLAS:2022exb}, and CMS~\cite{CMS:2018zza,CMS:2023lfu} Collaborations are also shown for comparison. } 
	\label{fig:RAA_I_0}%
\end{figure}

\begin{figure}[tb!]
	\centering 
	\includegraphics[width=0.45\textwidth]{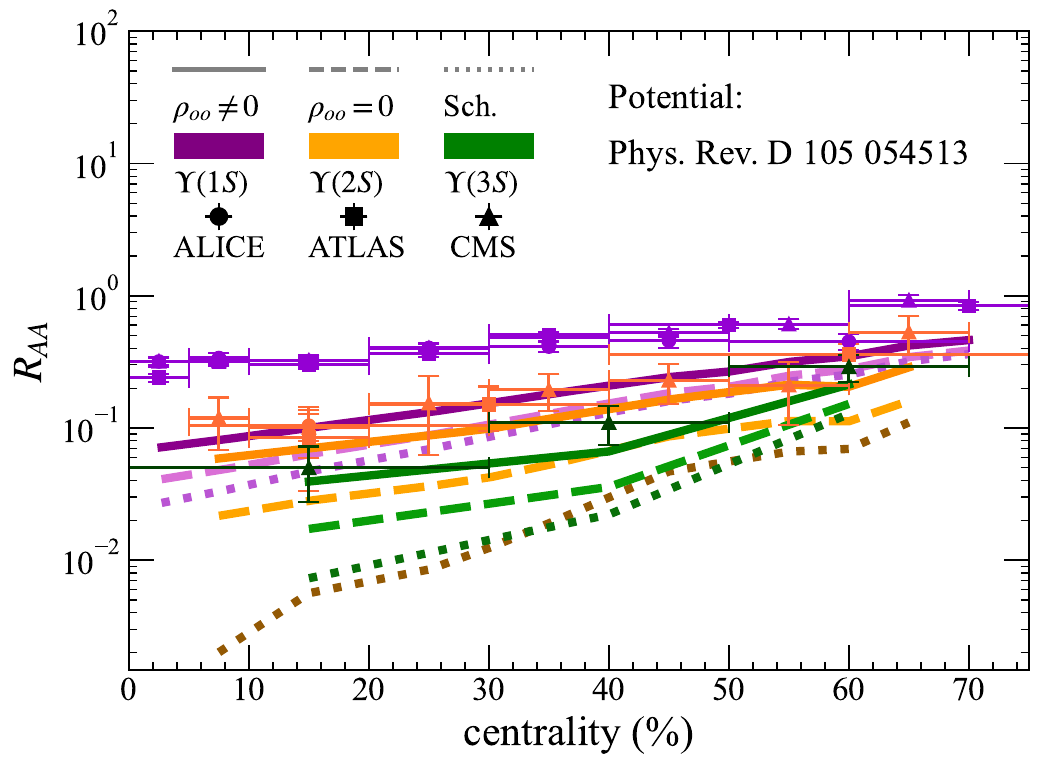}
    \includegraphics[width=0.45\textwidth]{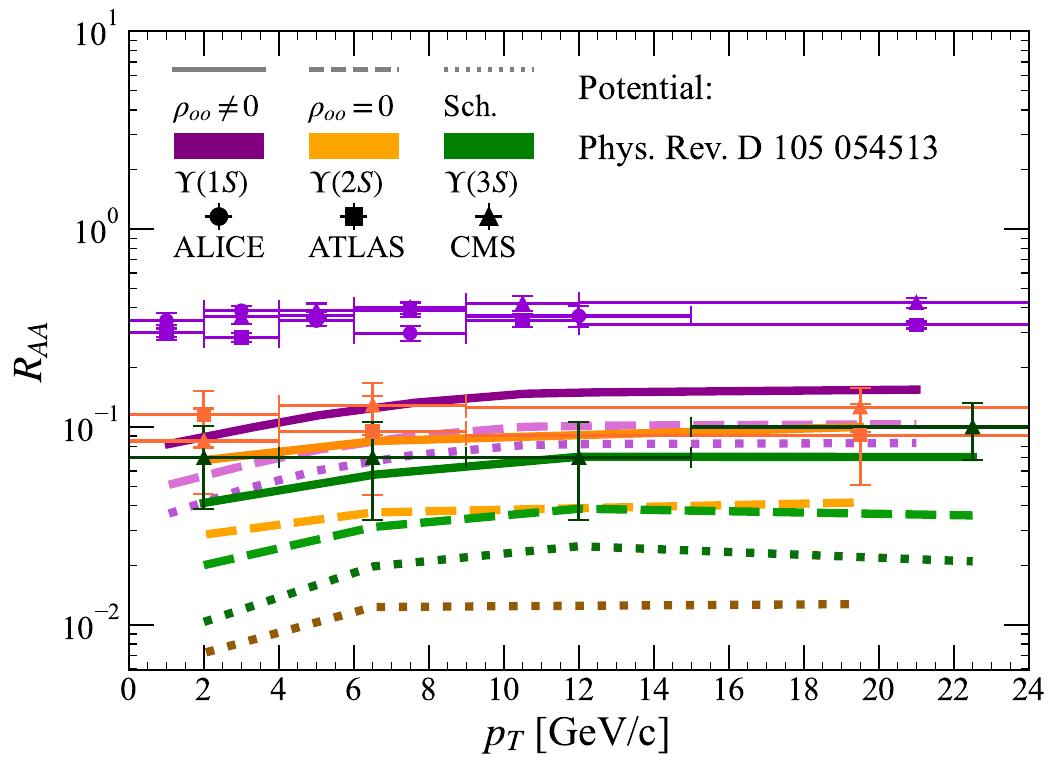}
	\caption{(Color online) Same as Fig.~\protect{\ref{fig:RAA_I_0}} but for the Burnier--Kaczmarek--Rothkopf potential~\cite{Burnier:2015tda}.} 
	\label{fig:RAA_I_1}%
\end{figure}

We describe the in-medium evolution of the QGP with $(2+1)$-dimensional \textsc{music} viscous hydrodynamic simulations~\cite{Schenke:2010nt,Schenke:2010rr} and obtain the local temperature $T =T(\mathbf{x}_0 + \mathbf{v}t,t)$ at the $b\bar{b}$ pair's center-of-mass position. When the local temperature falls below the dissociation threshold (switching temperature) $T_d = 0.16\;\mathrm{GeV}$ or the hydrodynamic background has ended, the evolution continues in vacuum with only the Cornell Hamiltonian, and all collapse operators are set to zero.
This, together with the remaining physical and numerical setup---including the Glauber initial condition, EPS09 nuclear parton distribution functions, Cronin effect, and the normalized initial transverse momentum distribution of the bottomonium bound states---follows the prescription of the authors' previous Schr\"odinger-equation-based analysis~\cite{Zheng:2025bdc}, to which the reader is referred for further details.

Figures~\ref{fig:RAA_I_0} and~\ref{fig:RAA_I_1} display the $R_{AA}$ obtained with the HotQCD and Burnier--Kaczmarek--Rothkopf potentials, respectively, whose parametrizations are detailed in \ref{app:B}. Solid and dashed lines respectively correspond to full evolution with pQCD motivated and singlet-only initial states, whereas dotted curves are for evolution without octet-to-singlet regeneration for comparison\footnote{We note that once octet-to-singlet regeneration is turned-off in the Lindblad framework, the quantum state evolution is analytically equivalent to the Schr\"odinger equation evolution of the singlet sector. We have also verified this numerically.}. The centrality dependence is shown for $p_T < 30\;\mathrm{GeV}/c$, and the $p_T$ dependence is integrated over the $0$--$80\%$ centrality class, as in the authors' previous analysis based on the Schr\"odinger framework~\cite{Zheng:2025bdc}. In both scenarios, $R_{AA}$ rises toward peripheral collisions and exhibits a plateau as a function of $p_T$.

We observe a non-negligible quantum regeneration effect---the difference between the solid and dotted curves, particularly for the excited states, where the Lindblad evolution yields visibly larger $R_{AA}$ than the calculation without regeneration. For the HotQCD potential (Fig.~\ref{fig:RAA_I_0}), the regeneration-induced difference decreases with increasing centrality, while its $p_T$ dependence shows a modest decline at low $p_T$ followed by a rise for $p_T \gtrsim 12\;\mathrm{GeV}/c$. Notably, for $\Upsilon(2S)$ and $\Upsilon(3S)$ the yields with quantum regeneration exceed those of the Schr\"odinger-based treatment by approximately one order of magnitude, demonstrating that regeneration dominates the production of excited bottomonia. For the ground state, the yields with quantum regeneration are approximately $16~(1.8)$ times those without regeneration in central (peripheral) collisions, indicating that even $\Upsilon(1S)$ benefits substantially from regeneration. These are qualitatively consistent with Ref.~\cite{Wu:2025lcj}, but the quantitative production rates remain lower than the equilibrium rate estimated therein. Additionally, different choices of initial conditions demonstrate that the effect of bottomonium suppression is overestimated with the singlet-only initial condition compared to the pQCD-motivated one, especially in the $p_T$-dependent calculation, where the yields are reduced by factors of 3, 4, and 3.3 for $\Upsilon(1S)$, $\Upsilon(2S)$, and $\Upsilon(3S)$, respectively. Sequential suppression~\cite{Digal:2001ue}---the stronger suppression of excited states relative to the ground state---becomes unapparent in both the centrality dependence and the transverse momentum $p_T$ dependence with $f_{oo} = 5/7$, while with the singlet-only initial condition it becomes apparent only in peripheral collisions (centrality $\gtrsim 60\%$).

Similar trends are observed for the Burnier--Kaczmarek--Rothkopf potential (Fig.~\ref{fig:RAA_I_1}): the difference between solid and dotted curves decreases with increasing centrality, particularly for the excited states. Sequential suppression, though still absent without quantum regeneration, begins to emerge when singlet-octet transitions are included, most visibly in the $p_T$-differential panel. The $R_{AA}$ value for the bound state is $2$--$3$ times larger than in the other case, indicating weaker in-medium suppression. However, the regeneration-induced enhancement is markedly smaller than that in Fig.~\ref{fig:RAA_I_0}, suggesting that quantum regeneration is no longer the dominant production mechanism in this case, where mutual transitions between octet and singlet states are suppressed due to the small $V_I$. Comparing the lower plots of Fig.~\ref{fig:RAA_I_0} and Fig.~\ref{fig:RAA_I_1} shows that the difference between the solid and dashed curves is less evident in the latter, with the aforementioned reduction factors decreasing to roughly 1.5, 2.4, and 2 for $\Upsilon(1S)$, $\Upsilon(2S)$, and $\Upsilon(3S)$, respectively.
In all settings, the regeneration contribution is not sufficient to resolve the tension in describing the LHC data~\cite{ALICE:2018wzm, ATLAS:2022exb, CMS:2018zza, CMS:2023lfu} with the lattice-QCD-based potential~\cite{Bala:2021fkm} observed in the Schr\"odinger framework of color-singlet wavefunctions~\cite{Chen:2024iil}.

\section{Summary and conclusions}
\label{Sec:summary}
In this study, we have investigated, within an open quantum system framework, the bottomonium regeneration and suppression in Pb-Pb collisions at $\sqrt{s_{NN}}=5.02$ TeV. The $b\bar{b}$ density matrix, comprising color-singlet and color-octet blocks, evolves according to a Lindblad master equation with interaction operators derived from pNRQCD interaction potential. Quantum regeneration of singlet states from octet configurations is governed by the collapse operator, whose structure is matched to the nonperturbative imaginary potential from lattice QCD. Two parametrizations of the in-medium heavy-quark potential---one with a temperature-independent and one with a temperature-dependent real part---both constrained by lattice QCD calculations, respectively by Burnier--Kaczmarek--Rothkopf~\cite{Burnier:2015tda} and the HotQCD collaboration~\cite{Bala:2021fkm}, have been employed.

We first examined the time evolution of the diagonal density matrix elements in a static QGP medium. 
Early-time asymptotic behavior recovers the corresponding Schr\"odinger equation of the color-singlet wavefunction, and the density matrix eventually approaches an equilibrium one with the ratio for diagonal elements close to that of the degeneracy factor.

We then compute the bottomonium production in realistic Pb-Pb collisions at $\sqrt{s_{NN}}=5.02$ TeV.
The results demonstrate that octet-to-singlet regeneration plays a substantial role in final-state bottomonium production. For the HotQCD potential, regeneration enhances the $\Upsilon(1S)$ yield by a factor of 1.8 to 16 relative to a Schr\"odinger-equation-based treatment in the $0-70\%$ centrality class, indicating that regeneration dominates the production of excited bottomonia. For the Burnier--Kaczmarek--Rothkopf potential, sequential suppression---the stronger suppression of excited states relative to the ground state---begins to emerge when octet-singlet transitions are included, most visibly in the $p_T$-differential $R_{AA}$, whereas it remains absent without regeneration. 

While the present calculations do not yet achieve full quantitative agreement with the experimental data for $R_{AA}$, the pronounced regeneration effects and the emergence of sequential suppression establish the necessity of a complete open quantum system treatment for bottomonium phenomenology. Our calculation also shows evident sensitivity to the initial condition of the density matrix, especially the ratio between color-singlet and octet states. This calls for a more rigorous calculation of the unquenched $b$-$\bar{b}$-pair density matrix from, e.g., a perturbative QCD approach.

\section*{Acknowledgements}
The authors thank Puyuan Bai, Jin Hu, Ziyi Liu, Yi Wang, and Wan Wu for helpful discussions. This work is supported by Tsinghua University under Grant Nos. 04200500123, 531205006, and 533305009. The authors also acknowledge the support of the High Performance Computing Center, Tsinghua University.

\appendix

\section{Bottomonium potential parametrizations based on Lattice QCD} \label{app:B}
Two parametrizations of the in-medium heavy-quark potential are employed in this work, which respectively fit two sets of lattice QCD calculations of the complex-valued in-medium heavy-quark potential~\cite{Burnier:2015tda, Bala:2021fkm}.

\textit{HotQCD potential.} The first is a direct fit~\cite{Chen:2024iil} to the static $b \bar{b}$ potential from the lattice NRQCD calculation of the distance-dependent imaginary-time correlation assuming Gaussian-shaped spectral functions at finite temperature, computed by the HotQCD collaboration~\cite{Bala:2021fkm}, yielding a Cornell-like form for the real part without explicit temperature dependence:
\begin{equation}
V_s^{\rm lQCD}(r) = \sigma_{\rm lQCD}\,r - \frac{\alpha_{\rm lQCD}}{r}, \label{eq:Vs_lQCD}
\end{equation}
\begin{equation}
V_o^{\rm lQCD}(r) = \frac{1}{8}\,\frac{\alpha_{\rm lQCD}}{r}, \label{eq:Vo_lQCD}
\end{equation}
with $\sigma_{\rm lQCD} = 0.22\;\mathrm{GeV}^2$ and $\alpha_{\rm lQCD} = 0.3805$. The prefactor $1/8 = 1/(N_c^2-1)$ in the octet potential is the color factor for single-gluon exchange in the color-octet channel.
For the imaginary part of this potential, the lattice QCD constrained parametrization is adopted in this work~\cite{Bala:2021fkm}
\begin{equation}\label{eq:VI_CornellLike}
    \frac{V_I(r;T)}{T} = (rT)^{1.2} + 0.54\,(rT),
\end{equation}

\textit{Burnier--Kaczmarek--Rothkopf potential.} The second parametrization incorporates temperature dependence via a Hard Thermal Loop (HTL) inspired model~\cite{Lafferty:2019jpr} that fits the complex-valued potential from Burnier, Kaczmarek, and Rothkopf's lattice QCD calculation with $2+1$ flavors of dynamical light quarks discretized with the asqtad action, with the spectral function extracted using the Bayesian reconstruction method~\cite{Burnier:2015tda}:
\begin{equation}
V_s^{\rm HTL}(r,T) = \frac{2\sigma}{m_D} - \frac{e^{-m_D r}(2+m_D r)\sigma}{m_D} - \tilde{\alpha}_s\Bigl[m_D + \frac{e^{-m_D r}}{r}\Bigr], \label{eq:Vs_HTL}
\end{equation}
\begin{equation}
V_o^{\rm HTL}(r,T) = \frac{1}{8}\,\tilde{\alpha}_s\Bigl[m_D + \frac{e^{-m_D r}}{r}\Bigr], \label{eq:Vo_HTL}
\end{equation}
with the vacuum string tension $\sigma = 0.2\;\mathrm{GeV}^2$ and $\tilde{\alpha}_s = 0.4105$. The Debye mass $m_D(T)$ encodes the screening of the color interaction by the thermal medium and is parametrized as a piecewise interpolation~\cite{Lafferty:2019jpr}:
\begin{align}\label{eq:mD}
    &m_D(T) = \notag \\
    &T \times
    \begin{cases}
        0, & T \le 0.126\;\mathrm{GeV},\\
        c_1 \ln(\alpha_1 T - \beta_1) - c_2 T, & 0.126 < T < 0.20\;\mathrm{GeV},\\
        c_3 \ln(\alpha_2 T - \beta_2) - c_4 T^{\gamma} + c_5, & T \ge 0.20\;\mathrm{GeV},
    \end{cases}
\end{align}
where the coefficients are $c_1 = 1.275$, $\alpha_1 = 170.8$, $\beta_1 = 19.41$, $c_2 = 7.593$, $c_3 = 0.6724$, $\alpha_2 = 9.535 \times 10^4$, $\beta_2 = 1.262 \times 10^4$, $\gamma = 0.12$, $c_4 = 7.864$, $c_5 = 2.497$.
For the Burnier--Kaczmarek--Rothkopf potential, the imaginary part derived from the generalized Gauss law~\cite{Lafferty:2019jpr} comprises a Coulomb term and a string term expressed through the Meijer $G$-function:
\begin{align}\label{eq:VI_HTL}
V_I^{\rm HTL}(r,T) = &\tilde{\alpha}_s T \, \phi(m_D r) + \notag \\
&\frac{\sqrt{\pi}}{4} m_D T \sigma \, r^3 \, G_{2,4}^{2,2}\!\left(\begin{matrix}-\frac{1}{2},-\frac{1}{2} \\ \frac{1}{2},\frac{1}{2},-\frac{3}{2},-1\end{matrix} \;\middle|\; \frac{1}{4}m_D^2 r^2\right),
\end{align}
where $\phi(x)$ is the dimensionless Coulomb spectral function and $G_{2,4}^{2,2}$ is the Meijer $G$-function arising from the Fourier transform of the HTL permittivity.

\bibliographystyle{elsarticle-num}
\bibliography{references}






\end{document}